\documentclass{article}
\usepackage{a4wide}

\begin{document}

\title{Steganography from weak cryptography}
\author{B. \v{S}kori\'{c}}
\date{}

\maketitle

\begin{abstract}
\noindent
We introduce a problem setting which we call ``the freedom fighters' problem''. It subtly differs from the prisoners' problem. 
We propose a steganographic method 
that allows Alice and Bob to fool Wendy the warden in this setting.
Their messages are hidden in encryption keys.
The recipient has no prior knowledge of these keys, and
has to cryptanalyze ciphertexts in order to recover them.
We show an example of the protocol and give a partial security analysis.
\end{abstract}

\section{Introduction}

Usually, the aim of steganography is to send a secret message $M$ hidden in
an ordinary data stream $S$ (the `covertext') in such a way that the adversary does not suspect the presence of~$M$.
This setting was formalized by Simmons in \cite{Sim83}, where he introduced the ``prisoners' problem''.
Two prisoners, called Alice and Bob, are located in different cells and wish to devise an escape plan.
They are allowed to exchange written messages.
However, Wendy the warden inspects all messages. If any message looks suspicious she will immediately punish them, ruining all their hopes of escape.
Her suspicion is aroused by 
e.g. references to escape, unusual phrases and formatting,
or anything resembling encryption.

In this paper we describe a steganographic technique for a somewhat different setting, which could be called the ``freedom fighters' problem''.
The freedom fighters Alice and Bob wish to plan an event.
They communicate over an insecure channel which is eavesdropped on by their powerful adversary Wendy.
The circumstances are as follows.
\begin{itemize}
\item 
Wendy allows Alice and Bob to discuss anything they wish. She never blocks a message.
\item 
Wendy will punish Alice and Bob if she catches them putting their plans into action.
\item 
Alice and Bob know that Wendy will watch them very closely if they use unbreakable\footnote{
This is the main difference with the prisoners' problem, where any form of crypto is punished.} 
crypto. Such surveillance will prevent them from realizing their event.
\item
Alice and Bob know that Wendy is highly adept at steganalysis and cryptanalysis.
\end{itemize}
The aim of the freedom fighters is to discuss their plan without Wendy learning what the plan is, and then to put the plan into action.
We propose the following solution.
Alice encrypts a covertext with a key that carries the secret mesage~$M$. She intentionally uses a weak cipher that is relatively easy to break, so that Bob can recover the key by breaking the encryption.
Wendy will of course also break the encryption, but she will 
focus far more on the covertext than on the
precise value of the employed key.
In order not to alert Wendy to the message present in the key, Alice and Bob use a second cipher to encrypt the hidden message so that the key looks random.

This scheme is different from \cite{Sim84,And96}, where messages are hidden in the random input of randomized signatures. 
Our system is effective because Wendy has no reason to believe that Alice and Bob are sending ciphertext to each other {\em for which the recipient does not have the decryption key}. 
Below we present an example of this scheme.

\section{An example of the protocol}
Alice and Bob have a shared secret key~$K$.
They have agreed on two symmetric ciphers:
A strong cipher ${\cal C}_1$ and a second, weak cipher ${\cal C}_2$.
The weak cipher is e.g. a block cipher with a known weakness such as a too short key, or an impressive-looking but flawed cipher cooked up by Alice and Bob.
${\cal C}_2$ works with keys of length~$\ell_2$.
Alice wants to send a secret message $M$ to Bob, hidden in covertext~$S$.
They perform the following steps:
\renewcommand{\labelenumi}{\arabic{enumi}.}
\begin{enumerate}
\item
Alice encrypts the message $M$ with the shared key $K$.
\[
	\kappa=E_{K}^{(1)}[M].
\]
The superscript `1' refers to the cipher ${\cal C}_1$.
She then cuts the ciphertext into $n$ pieces of length $\ell_2$
(padding if necessary),
\[
	\kappa = \kappa_1|| \cdots    ||  \kappa_n.
\]
\item
Alice composes a number of covertext messages $S_i$, $i=1,\ldots,n$.
The length of these messages is arbitrary. The messages are written in ordinary
language, or in some other highly redundant format.
\item
Alice computes the following encryptions, using the weak cipher:
\[
	c_i = E^{(2)}_{\kappa_i}[S_i].
\]
Over time she sends these $n$ ciphertexts to Bob.
\item
For each $i$,
Bob {\em breaks} the encryption, obtaining $S_i$ and $\kappa_i$ from~$c_i$.
His ability to do so is guaranteed by the weakness of the cipher 
${\cal C}_2$ and the fact that $S_i$ contains a lot of redundancy\footnote{
To aid Bob's cryptanalysis, Alice may also include pieces of plaintext in $S_i$ that are known to Bob beforehand.
}.
\item
Using the shared key $K$, Bob reconstructs Alice's message as follows 
\[
	M = D_K^{(1)}\left[\kappa_1|| \cdots    ||  \kappa_n
	\vphantom{M^M}\right].
\]

\end{enumerate}
\noindent
Then Bob sends a reply to Alice in the same way, etc.
(Alternatively, Alice and Bob exchange ciphertexts in an interleaved manner, i.e. Alice sends $c_1^{\rm Alice}$, then Bob sends $c_1^{\rm Bob}$ in response, then Alice sends $c_2^{\rm Alice}$ etc.
This allows for more natural looking covertexts, since communication with one message at a time looks more natural than $n$ messages in one direction followed by $n$ messages in the other direction. 
The exchange of covertexts looks like a normal conversation, with each covertext containing references to the preceding covertexts.
The message $M$, on the other hand, can of course be reconstructed only after receiving $n$ covertexts; so in terms of message exchanges
the hidden conversation is $n$ times slower than the cover conversation.)

Of course, Wendy too is capable of breaking the ${\cal C}_2$ encryption.
From the ciphertexts $c_i$ she obtains $S_i$ and $\kappa_i$.
She will assume that the $S_i$ are genuine messages.
{\em She has no reason to assume that Alice is sending encrypted data for which Bob has no decryption key!}
In contrast, it is entirely believable to Wendy that they are entrusting
their secrets to a weak cipher. There are many historical examples of people thinking 
that their home-brewed ciphers are invincible.

Being a good cryptanalyst,
Wendy will wonder what kind of key schedule Alice and Bob are using.
She will notice that in general $\kappa_1\neq\kappa_2\cdots\neq\kappa_n$. However, the 
cipher ${\cal C}_1$ will prevent her from finding any regularity,
provided that ${\cal C}_1$ is strong enough.
The lack of regularity in the sequence of keys $\kappa_1,\cdots,\kappa_n$ will not arouse 
suspicion in Wendy; 
Alice and Bob may have set up a list of one-time keys in the past, or they
could be using some key updating schedule. 

Alice and Bob make sure that the exchange of covertext messages looks 
`normal', which in this case means that it must look like an exchange of highly confidential information, 
i.e. the kind of data that would never be sent in plaintext. 
They may also sometimes refer to a `key schedule' for determining the
$\kappa_i$ keys (of course completely fake), thus convincing Wendy that the $\kappa_i$ values are of no 
direct importance.
A nice property of our scheme is that 
a convincing-looking covertext can misdirect Wendy in many ways. 
Most notably, the covertext may directly contradict the secret message.

{\it Remark}: 
The bit rate of the hidden channel is rather low: $\ell_2$ bits of ciphertext per
exchanged message.

\section{Security analysis}

Of course, publishing about a steganographic scheme gives it away.
Once the adversary suspects that there could be a payload in the encryption keys
$\kappa_i$,
she will start paying attention to them and start distrusting~$S_i$.
The steganalysis is now a matter of 
\begin{enumerate}
\item[A.]
detecting if there is anything fishy about the set $\{\kappa_i\}$,
and 
\item[B.]
breaking the ${\cal C}_1$-encryption.
\end{enumerate}
A thorough analysis of part A is nontrivial, and we will not attempt it in this paper.
The fact that the encryption key varies does not, in itself, automatically raise suspicion. 
First, 
Alice and Bob may simply have agreed on a list of one-time keys.
Second, they may be using a protocol involving session key updates.
Several protocols are known in the literature where a session key gets updated, e.g. using a hash chain, in order to provide backward security.
It depends on the circumstances if Wendy has reasons to disbelieve these possibilities.
(Note that this has implications for so-called `deniable encryption'.)

The difficulty of part B of the steganalysis 
directly translates to the difficulty of cryptanalysis.
The message $M$ remains hidden from Wendy if the cipher ${\cal C}_1$ is strong enough. 

{\it Remark}: If Wendy succeeds in part A, then, in the freedom fighters' problem setting, Alice and Bob have lost, even though $M$ remains safe.
They have become suspicious and are put under surveillance.

\subsection*{Acknowledgements}
We thank Stefan Katzenbeisser and Klaus Kursawe for useful suggestions.

\bibliographystyle{plain}

\bibliography{stego}

\end{document}